\documentclass[twocolumn,prl,floatfix]{revtex4}
\usepackage{hyperref}
\usepackage{graphicx}
\usepackage{amsfonts}
\usepackage{amssymb}
\usepackage{amsmath}
\usepackage{mathrsfs}
\usepackage{commath}
\usepackage{comment}
\usepackage{dsfont}

\usepackage{txfonts}
\usepackage{lipsum}
\usepackage{color}
\usepackage{wasysym}

\usepackage{gensymb}
\usepackage{tabularx,ragged2e,booktabs,caption}
\usepackage{subfigure}
\usepackage{tabularx,capt-of}
\usepackage{multirow}

\usepackage{makecell,booktabs}

\definecolor{dark-green}{RGB}{0, 128, 0}

\definecolor{red}{RGB}{200, 0, 0}

\usepackage[utf8]{inputenc}

\usepackage{array}
\usepackage{braket}

\newcommand{\dd}{{\rm d}}
\newcommand{\eff}{{\rm eff}}

\DeclareMathOperator{\Tr}{Tr}

\usepackage{graphicx}

\begin{document}
\title{Nonlinear edge modes from topological 1D lattices}

\author{Lucien Jezequel}
\affiliation{Univ Lyon, ENS de Lyon, CNRS, Laboratoire de Physique, F-69342 Lyon, France}
\author{Pierre Delplace}
\affiliation{Univ Lyon, ENS de Lyon, CNRS, Laboratoire de Physique, F-69342 Lyon, France}

\begin{abstract}
We propose a method to address the existence of topological edge modes in one-dimensional (1D) nonlinear lattices, by deforming the edge modes of linearized models into solutions of the fully nonlinear system. For large enough nonlinearites, the energy of the modified edge modes may eventually shift out of the gap, leading to their disappearance. We identify a class of nonlinearities satisfying a generalised chiral symmetry where this mechanism is forbidden, and the nonlinear edge states are protected by a topological order parameter. Different behaviours of the edge modes are then found and explained by the interplay between the nature of the nonlinarities and the topology of the linearized models.
\end{abstract}
\maketitle

Since the discovery of the quantum hall effect \cite{Klitzing}, the number of physical systems exhibiting topological boundary modes has been constantly increasing. Originally found in condensed matter, robust edge states are now found in virtually all wave systems \cite{lutopological2014,Delplace1075,Nash14495} showing the ubiquity of topological edge modes independently of their physical implementation. 
The robustness of the edge modes is understood through the celebrated bulk-boundary correspondence that relates the number of edge modes to a topological invariant of the bulk bands as long as the spectrum is gaped \cite{HatsugaiPRL1993,HatsugaiPRB1993,grafbulkedge2018,ProdanEmilSchultz}.

From the beginning, the theory of topological edge modes has been tied to linear systems concepts such as eigenmodes and  energy spectra.
Actually, many platforms used to implement those topological properties, such as polaritons and photonic lattices, fluids, networks of springs and electric circuits, also naturally exhibit nonlinear behaviours.  This recently stimulated a growing interest in the interplay of topology with nonlinearities \cite{Secli2019,Xia2020,NonlinearReview},  with applications to topological lasing \cite{TopologicalLaser,Gundogdu20} and topological synchronisation \cite{sone2020topological}.  
Regarding the edge states in nonlinear systems, investigations in specific cases showed the existence of nonlinear bulk and edge solitons  with more or less similarities with the linear case \cite{LeykamNonLinEdgeSoliton,ma2020existence,Nonlintopo,PhononTopo, tempelman2021topological,zhou2021topological,Engelhardt_2017,PhysRevB.102.115411}. In some examples in 1D, the energy of the edge modes was found to depend of the amplitude \cite{PhysRevLett121163901,PhysRevA100063830,ma2020existence,tempelman2021topological, ezawa2021topological}, while in others, the energy was found to be fixed \cite{Nonlintopo,hadad2018self,PhononTopo}. These results suggest that the concept of stationary topological edge mode seems generalizable to the nonlinear realm, with however a lack of a systematic theoretical understanding that goes beyond the scope of sporadic examples.

In this work, we provide criteria for the existence of nonlinear topological edge states. We focus on nonlinear Schr\"odinger equations on 1D lattices, whose nonlinear Hamiltonian $H_\psi$ splits into a linear topological part $H_{\text{topo}}$ and a nonlinear one $H_{\psi,\text{NL}}$ as
\begin{equation}
   i\partial_t \ket{\psi} = H_\psi \ket{\psi} = ( H_{\text{topo}}+H_{\psi,\text{NL}}) \ket{\psi} \ .
    \label{eq:Equationdebase}
\end{equation} 
 The burning question to ask is then: What are the conditions for the edge states of the linear topological model $H_{\text{topo}}$ to survive the presence of nonlinearities $H_{\psi,\text{NL}}$? 
To answer this question, we propose a method based on exact perturbation theory that generates the edge modes and their energy of  nonlinear systems of the form \eqref{eq:Equationdebase}. The idea is to start with the edge mode of the linearised system at small amplitude and then increase smoothly the amplitude of the mode. As the relative strength of the nonlinearities increases, we are then able to deform the initial linear edge mode in a way that it remains a stationary edge solution of the nonlinear dynamics. The method can eventually reach nonlinear edge modes with a quite high amplitude as long as their energy remains in the spectral gap of the  linearized dynamics. If this condition stops being fulfilled, the nonlinear edge state is then quickly delocalised into the bulk \cite{tempelman2021topological} and become unstable \cite{Chaunsalistability}. 
Then, we extend the notion of chiral symmetry to nonlinear systems, and show that chiral symmetric nonlinearities prevent such a delocalisation, thus protecting the nonlinear edge state. We then characterize those robust nonlinear edge modes with a local topological index. Our theory is illustrated with two nonlinear generalisations of the SSH model -- one being chiral and one which is not -- and confirmed numerically.

A concrete situation where the nonlinear Schr\"odinger equation modifies a 1D topological lattice model, is that of an SSH chain with couplings $t_1$ and $t_2$ between nearest neighbors and a on-site Kerr-like nonlinearity\cite{NonlinearReview,ma2020existence} similar to those appearing in the Gross-Pitaevskii equation for Bose-Einstein condensates \cite{BoseEinstein}:
\begin{equation}
\left\{ 
\begin{aligned}
    &i \partial_ta_j = t_1b_j + t_2b_{j-1}+ |a_j|^2a_j\\
    &i \partial_tb_j = t_1a_j + t_2a_{j+1}+ |b_j|^2b_j \ . 
\end{aligned}  \right. \label{eq:SSHnonchiral}
\end{equation}
A state $\ket{\psi}$ of the system can be decomposed in the basis of the two sublattices as
\begin{align}
    \ket{\psi} = \begin{pmatrix} \ket{\psi_A} \\\ket{\psi_B}    \end{pmatrix} =   \begin{pmatrix}\sum_j a_j\ket{j,A}\\\sum_j b_j\ket{j,B}\end{pmatrix}
\end{align}
 where $j$ labels the unit cell. The application of $H_\psi$ to such a state then gives the vector $H_\psi\ket{\psi}$ which expands as
\begin{align}
H_\psi \ket{\psi} = \begin{pmatrix}
 \sum_j\left(t_1b_j + t_2b_{j-1}+ |a_j|^2a_j \right)\ket{j,A}\\
\sum_j\left(t_1a_j + t_2a_{j+1}+ |b_j|^2b_j\right)\ket{j,B} 
\end{pmatrix} \ .
\label{eq:HpsiKerr}
\end{align}
for the model \eqref{eq:SSHnonchiral}.
The linear SSH model is recovered when $\ket{\psi}$  is small in amplitude. In that case, this model is known to have a gap in energy around $E=0$ when $|t_1|<|t_2|$, except for stationary topological edge modes which are localised at each end of the chain \cite{Zakphase,Hatsugai}. We then would like to know how these edge modes survive the introduction of the nonlinearities, such as in \eqref{eq:HpsiKerr}. We show below that edge states exist in nonlinear Schr\"odinger models when three conditions are met:
\begin{itemize}
    \item (i) The linearised model has an edge state which is in the gap of the bulk bands.
    \item (ii) The differential of $H_\psi$ around any state $\ket{\psi}$ is Hermitian, i.e. $\dd H_\psi^\dagger = \dd H_\psi$.
    \item (iii) The nonlinear Hamiltonian $H_\psi$ verifies the $U(1)$-symmetry  $H_{e^{i\phi}\psi}\left(e^{i\phi}\ket{\psi}\right)=e^{i\phi}H_\psi \ket{\psi}$ for all $\ket{\psi}$.
\end{itemize}

Assumption (i) is quite natural, as we search nonlinear edge states as resulting from the deformation of linear ones. Assumption (ii) is made to guarantee that the energy $E$ of the state remains real.
And assumption (iii) is needed to insure that finding non-linear states of real energy $ E\ket{\psi}=H_\psi \ket{\psi}$ also generates solutions of \eqref{eq:Equationdebase} of the form $\ket{\psi(t)} =\ket{\psi}e^{-iEt}$. Note that our model \eqref{eq:SSHnonchiral} satisfies these three conditions. Note that those general hypothesis do not depend on the order of the nonlinear terms. Our method is thus not specific to Kerr-like terms and applies to arbitrary nonlinearities as long as the three hypothesis above are satisfied.

We now provide an explicit method to construct the nonlinear edge modes assuming that the three conditions above are met. For that let us study the space of edge states $\ket{\psi}$ of energy $E$ of $H_\psi$ that we define as being spanned by the doublet $(E,\ket{\psi})$. The key idea to explore this space is to parameterize it with a continuous parameter $s$ and to derive the evolution equation for the states close in $s$. If $(E_s,\ket{\psi_s})$ is a doublet such that for all $s$, $\ket{\psi_s}$ is a solution of:
\begin{equation}
    E_s\ket{\psi_s} = H_\psi \ket{\psi_s} \, 
    \label{eq:stationnary}
\end{equation}
then, by differentiating this equation along $s$, one finds that the condition for this path to exist is to satisfy the following evolution equation:
\begin{equation}
\label{eq:masterequation}   
\left(\dd H_{\psi_s}-E_s\right) \ket{\partial_s \psi_s} = (\partial_s E_s) \ket{\psi_s} 
\end{equation}
where the differential $\dd H_\psi$ is a linear operator that describes the variation of $H_\psi$ for a small perturbation $\ket{\partial_s \psi}$ around $\ket{\psi}$. One can thus interpret it as an effective Hamiltonian $H_{\eff,s}\equiv \dd H_{\psi_s}$ when linearizing around $\ket{\psi_s}$.

The free variables of \eqref{eq:masterequation} are $\ket{\partial_s\psi_s}$ and $\partial_sE_s$. It follows that for a lattice with $n$ sites, \eqref{eq:masterequation} is a system of $n$ differential equations with $n+1$ variables. Thus, at fixed $s$, the space of solutions $(\partial_sE_s,\ket{\partial_s\psi_s})$ of \eqref{eq:masterequation} is a vector space of dimension at least one. Therefore, there exists (at least) a solution to our evolution equation. Moreover, hypothesis (ii) implies that there is solution with $\partial_sE_s$ real. One can therefore reconstruct from \eqref{eq:masterequation} the continuous family of stationary solutions $(E_s,\ket{\psi_s})$ of \eqref{eq:stationnary} along this path, given an initial condition at $s=0$, that we choose to be $(E_{s=0},\ket{\psi_{s=0}})=(0,0)$. 

For an infinitesimal deviation away from this initial condition, $H_\psi$ can be linearised as $H_{\eff,0}=\dd H_0$. If this linear model hosts at least one edge mode of zero-energy $\ket{\psi}$, like in the SSH model, then $((\partial_sE_s)_{s=0} = 0,\ket{\partial_s\psi_s}_{s=0}=\ket{\psi})$ is a valid solution of \eqref{eq:masterequation} for $s=0$. Solving this differential system, one can therefore generate nonlinear edge states $\ket{\psi_s}$ with a growing amplitude as $s$ increases. If the linear model hosts multiple zero-energy edge modes as the SSH model (one on each edge), all of them could be used for the dynamic, leading to different non-linear edge modes.

So far, we have obtained the existence of solutions $(\partial_sE_s,\ket{\partial_s\psi_s})$ for \eqref{eq:stationnary} but  we have not shown yet that they remain localised near the edge. The question is the following: If the linear model at $s=0$ has an edge mode, is $\ket{\psi_s}$ also localised near the edge for $s>0$ ? In systems where coupling constants between sites decay quickly with their distance as in our illustrative nonlinear SSH model \eqref{eq:SSHnonchiral}, the answer is given by the Combes-Thomas theorem \cite{CombesThomas,AizenmanRandOperator}.
This theorem states that solutions $\ket{\partial_s\psi_s}$ of \eqref{eq:masterequation} are localised around $\ket{\psi_s}$ as long as $E_s$ lies in the bulk gap of $H_{\eff,s}$. If this condition stops being satisfied, $\ket{\psi_s}$ can quickly be delocalised as $\ket{\partial_s\psi_s}$  starts to strongly resonate with the nearby bulk modes.

\begin{figure}
    \centering
    \includegraphics[height=10cm,width = 8.5cm]{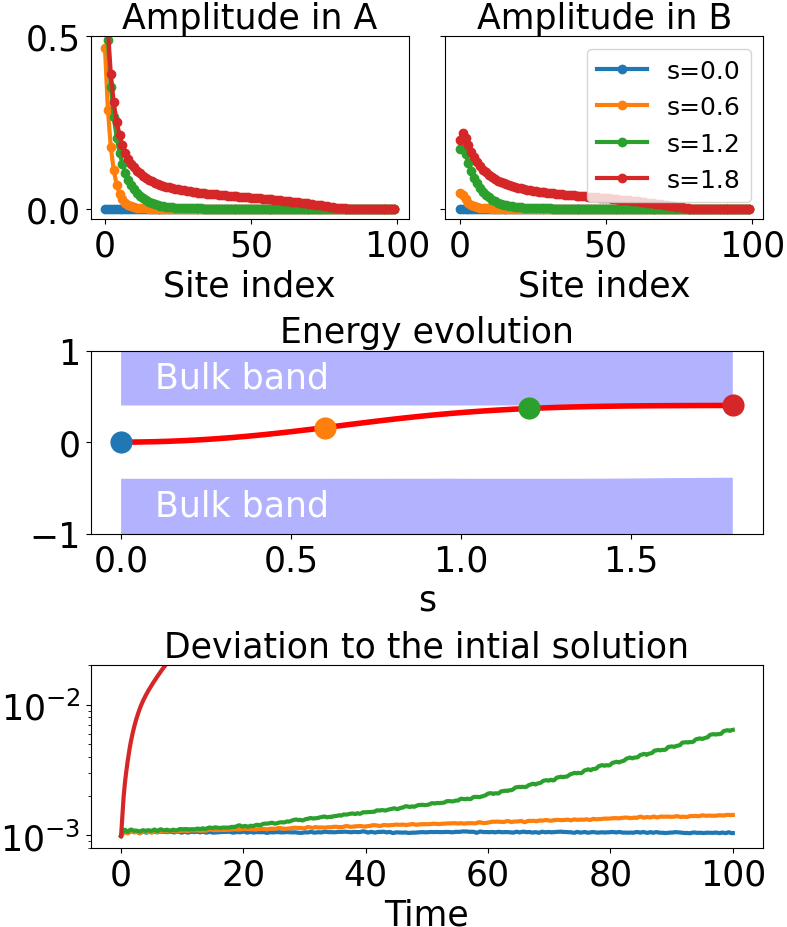}
    \caption{Numerical resolution of  \eqref{eq:masterequation} for the left edge modes of model \eqref{eq:SSHnonchiral}. We work with $100$ pairs of sites and the constants $t_1=0.6$ and $t_2=1$. The amplitude of $\psi_s$ is given on the sites of type A (up left) and B (up right) for different $s$. (center) The evolution of the energy $E_s$ of $\psi_s$ is drawn in red and the bulk bands of $H_{\eff,s}$ in light blue. (bottom) Deviation $\| \ket{\psi(t)}- \ket{\psi_s}\|$ between the stationary solution $\ket{\psi_s}$ and an initially perturbed one by a random vector of norm $10^{-3}$, for different values of $s$. }
    \label{figurenonchirale}
\end{figure}

In most cases, the system \eqref{eq:masterequation} must be solved numerically, using standard algorithmic methods \cite{RK4}. In particular we can solve this system for the Kerr-like nonlinear SSH model \eqref{eq:SSHnonchiral} as an illustration. As this is a model which verifies the general hypothesis (i-iii) we can therefore generate left edge states with a growing amplitude as $s$ increases (see Fig \ref{figurenonchirale}). For small amplitude, their shape remains close to the exponential shape of the edge states of the linearized model. But as the relative strength of the nonlinearities increase, the nonlinear edge states become more and more deformed. In particular, we observe that the nonlinear edge states becomes less localised as their energy $E_s$ approaches the bulk bands of $H_{\eff,s}$. Around $s\sim 1.5$ the energy touch those bands and the edge state becomes strongly delocalised. Therefore, the edge state is not topologically protected in the strong amplitude regime. Moreover, since the system is nonlinear, one can ask about the stability of such stationary solutions under small perturbations \cite{Panda,tempelman2021topological,Chaunsalistability}. In order to do so, we add a random perturbation $\ket{\delta \psi}$ to a stationary solution $\ket{\psi_s}$. Then, for the initial condition $\ket{\psi(t=0)}=\ket{\psi_s}+\ket{\delta \psi}$, we evaluate how far the solution $\ket{\psi(t)}$ of \eqref{eq:Equationdebase} deviates from the original stationary solution $\ket{\psi_s}$ by measuring the time evolution of their distance $\|\ket{\psi(t)}-\ket{\psi_s}\| = \sum_j |\psi_j(t)-\psi_{s,j}|$,  where $\psi_i$ denotes the $i$-site amplitude. If the deviation grows exponentially with time, the state is unstable. We find (see figure \ref{figurenonchirale}) that the edge state is stable as long as the   energy $E_s$ does not enter the bulk bands, which occurs at about $s\sim 1.2 -1.5$. Beyond this threshold, the states $\ket{\psi_s}$ become strongly unstable, highlighting again the criticality of this regime.

In practical situations, one would like to prevent this band touching to occur by constraining the energy at zero, in the middle of the gap of $H_{\text{eff},s}$. In 1D insulators, this protection role is made by the presence of a chiral symmetry. To follow that spirit, we introduce a generalisation of the chiral symmetry to nonlinear systems. This allows us to identify nonlinearities that forbid the energy-shift and therefore host edge states that are robust and topologically protected. Besides, unlike the general case discussed so far, the result does not require $H_\psi$ to satisfy a $U(1)$-symmetry, nor $\dd H_{\psi}$ to be Hermitian.

We say that a nonlinear operator $H_\psi$ satisfies a chiral symmetry if there is a bi-partition $A$ and $B$ of the degrees of freedom -- e.g. two sublattices -- such that
the state $H_\psi \ket{\psi}$ decomposes onto a single sublattice (say $B$) \textit{if}  $\ket{\psi}$ decomposes onto the other sublattice (say $A$). Put formally, one wants
\begin{equation}
    \ket{\psi}_B=0 \implies (H_\psi\ket{\psi})_A=0  \qquad  \text{(same for $A \leftrightarrow B$)} \ .
    \label{eq:Chiralcondition}
\end{equation}
In the linear case, \eqref{eq:Chiralcondition} imposes $H$ to be off-diagonal when written in the $A$ and $B$ basis. Our definition thus generalises the chiral symmetry. 

For the sake of concreteness, let us illustrate whether this symmetry is satisfied for a few different nonlinear terms.
For the Kerr nonlinearity of the model \eqref{eq:SSHnonchiral} we have $H_{\text{Kerr},\psi}\ket{\psi}= \sum_j a_j^3\ket{j,A}+b^3_j\ket{j,B}$, so $(H_{\text{Kerr,}\psi}\ket{\psi})_A=\sum_j a_j^3 \ket{j,A}\neq 0$ even when  $b_j=0$. So the Kerr nonlinearity is not chiral symmetric. Instead, we can introduce the Kerr \textit{inter-site} nonlinearity, of the form $H_{\text{inter-Kerr,}\psi}\ket{\psi} = \sum_j b_j^3\ket{j,A}+a^3_j\ket{j,B}$ that verifies $(H_{\text{inter-Kerr,}\psi}\ket{\psi})_A=\sum_j b_j^3 \ket{j,A}=0$ when $\ket{\psi}_B=0$ (meaning $b_j=0\, \forall j$), and same for $A \leftrightarrow B$. This nonlinearity is thus chiral symmetric. As additional examples, we give below a list of nonlinear terms with general exponents $\alpha$ or $\beta$ and classify them according to the chirality condition. It should be noted that this list stays valid if one does the exchange $A \leftrightarrow B$ everywhere. 
\vspace{0.2cm}

\begin{tabular}{ |l||c | c |c|c| }
\hline
$H_\psi\ket{\psi}$ & $a_j^\alpha\ket{j,A}$ & $b_j^\alpha\ket{j,A}$ & $a^\alpha_{j+1}b^\beta_{j}\ket{j,A}$  &  $b^\alpha_{j-1}b^\beta_{j+1}\ket{j,A}$\tabularnewline
   \hline
Chiral & No & Yes  & Yes & Yes \tabularnewline
   \hline
 \end{tabular}
\vspace{0.2cm}

Importantly, when $H_\psi$ is chiral symmetric, an initial linear edge state $\ket{\psi_{s=0}}$ living on a given sublattice can evolve through \eqref{eq:masterequation} as a stationary solution $\ket{\psi_{s>0}}$ that remains on the same sublattice.
Indeed if $\ket{\psi_s}$ is a solution of \eqref{eq:masterequation} satisfying $\ket{\psi_0} = 0$ and $\ket{\psi_s}_B = 0$, $\forall s$, then by writing \eqref{eq:masterequation} by blocks while assuming $E_s = 0$, one obtains
\begin{align}
\begin{pmatrix}H_{\text{eff,s}}^{AA} & H_{\text{eff,s}}^{AB}\\ H_{\text{eff,s}}^{BA}& H_{\text{eff,s}}^{BB}\end{pmatrix} \begin{pmatrix}\ket{\partial_s\psi_s}_A\\0\end{pmatrix} = 0 \ .
\label{eq:systchiral}
\end{align}

\begin{align}
    \ket{\psi} = \begin{pmatrix} \ket{\psi_A} \\0   \end{pmatrix} 
\end{align}

But differentiating the condition \eqref{eq:Chiralcondition} along the variable $\ket{\psi}_A$ yields $H^{AA}_{\text{eff},s} = 0$. So \eqref{eq:systchiral} reduces to
\begin{equation}
    H_{\text{eff,s}}^{BA}\ket{\partial_s\psi_s}_A = 0 \ .
    \label{eq:Masterequationchiral}
\end{equation}
\begin{figure*}[ht]
    \centering
    \includegraphics[width = 17cm,height=8cm]{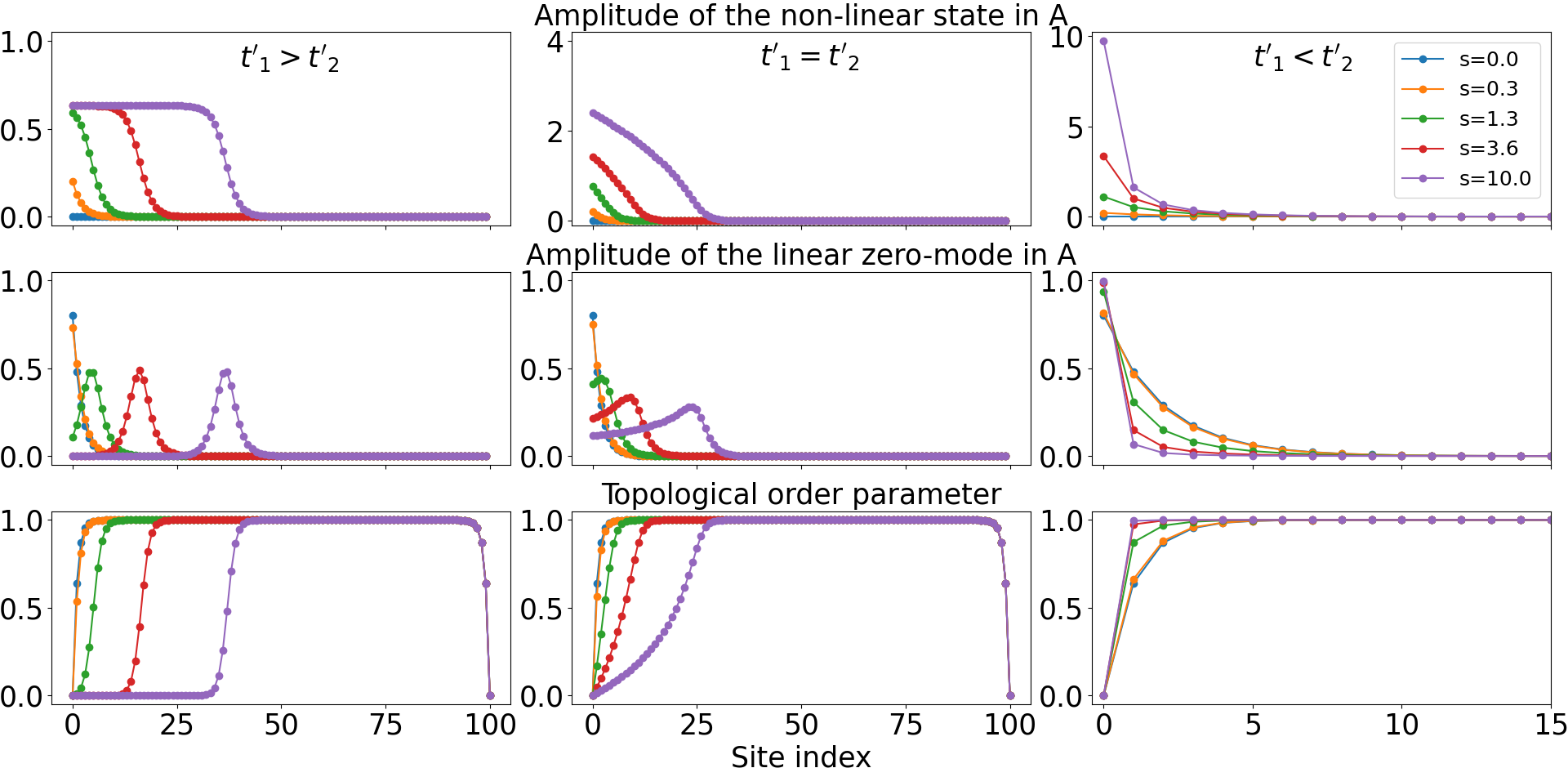}
    \caption{Numerical resolution of \eqref{eq:Masterequationchiral} for the left edge modes of the model \eqref{eq:SSHchiral}. We work with $100$ pairs of sites and $t_1=0.6$ $t_2=1$ everywhere. For the nonlinear couplings we take (left) $t_1'=1$ $t_2'=0$ (center) $t_1'=1$ $t_2'=1$ (right) $t_1'=0$ $t_2'=1$. We draw (up) the amplitude of $\psi_s$ on the A-sites (center) the zero-energy state of $H^{AB}_{\text{eff,s}}$ (down) the topological order parameter $I(x)$ where we took $\epsilon=\frac{1}{100}$}
    \label{figurechirale}
\end{figure*}

In order to know whether $H_{\text{eff,s}}^{BA}$ has localised zero-modes for each $s$, we use the theory of topological indices. For that purpose, we define the following operators in the spirit of super-symmetric approaches \cite{SupersymmetryVitelli,kane_topological_2014,PhysRevResearchSuperSymmetry}
\begin{equation}
    H'_{\text{eff,s}}= \begin{pmatrix}
    0& H^{BA}_{\text{eff,s}}\\ H^{BA \hspace{0.05cm} \dagger}_{\text{eff,s}}&0
    \end{pmatrix} \hspace{0.5cm} C= \begin{pmatrix}
    \mathds{1}_A& 0\\0&-\mathds{1}_B
    \end{pmatrix}
\end{equation}
\begin{equation}
    C= \begin{pmatrix}
    \mathds{1}_A& 0\\0&-\mathds{1}_B
    \end{pmatrix} \hspace{0.5cm} H= \begin{pmatrix}
    0& H^{BA}\\ H^{BA \dagger}&0
    \end{pmatrix} 
\end{equation}
so that $\{H'_{\text{eff,s}},C\}=0$ implying that $H'_{\text{eff,s}}$ is a chiral Hermitian operator associated to $H^{BA}_{\text{eff,s}}$. Next, we introduce the operator $P= \tanh(H'_{\text{eff,s}}/\epsilon)$
whose eigenstates are those of $H'_{\text{eff,s}}$, but whose spectrum $\tanh(E/\epsilon)$ flattens the bulk bands of $H'_{\text{eff,s}}$ and separates them from the zero energy edge states as the parameter $\epsilon \rightarrow 0$.
In practice, we want $\epsilon$ to be smaller than the bulk gap of $H'_{\text{eff,s}}$, but larger than the energy of the edge states, which is never rigorously zero in finite systems. Introducing then the step-function $\theta_{j}(j')$  which is 1 for $j' \geq j$ and 0 otherwise, we can finally define from $H'_{\text{eff,s}}$ the following \textit{topological order parameter} 
\begin{equation}
    \mathcal{I}(j) = \frac{1}{2}\Tr\left(C\left[\hat{\theta}_{j},P\right]P\right) 
\end{equation}
where $\hat{\theta}_{j}$ is the diagonal operator associated to $\theta_{j}$.
Behind its abstract definition, this local quantity is very useful. It can be shown to be a constant integer (related to the winding number in the periodic case \cite{noncommutativeindextheorem}) in regions where $H'_{\text{eff,s}}$ has no zero-modes and can only change when crossing regions with zeros modes of $H^{BA}_{\text{eff,s}}$ or $H^{BA,\hspace{0.05cm} \dagger}_{\text{eff,s}}$. In particular, there is a correspondence connecting the index variation $\Delta\mathcal{I}=\mathcal{I}(j_2)-\mathcal{I}(j_1)$ to the number of zero modes of $H^{BA}_{\text{eff,s}}$ localised in the interval $j_1 \leq j \leq j_2$ minus those of $H^{BA,\hspace{0.05cm} \dagger}_{\text{eff,s}}$ \cite{grafbulkedge2018,JezequelDelplace}. In particular, when $\Delta\mathcal{I}>0$, this correspondence implies that $H^{BA}_{\text{eff,s}}$ has at least $\Delta\mathcal{I}$ zero modes localised between $j_2$ and $j_1$.

If we take $j_1=0$ we can prove that $\mathcal{I}(j_1)=0$ as $\hat{\theta}_{j=0}= \mathds{1}$. Moreover, as long as the edge state do not invade the whole bulk, we have that $H_{\text{eff,s}} \approx H_{\text{eff,0}}$ far from the edges. So if we take $j_2$ far enough from the edges, then $\mathcal{I}(j_2)$ is just the index one would obtain in the bulk of the linearised model at small amplitude. Thus if we denote $\mathcal{I}$ this topological number, we see that $H^{BA}_{\text{eff,s}}$ is constrained to have at least $\mathcal{I}$ zero-modes localised on the left part of the chain. If $\ket{\psi_s}$ is a nonlinear edge mode it  thus implies that we have at least $\mathcal{I}$ possible choices for $\ket{\partial_s\psi_s}$ which are localised and verify \eqref{eq:Masterequationchiral}.

We now apply our nonlinear chiral theory to a concrete model that we solve numerically. As mentioned above, inter-sites Kerr nonlinearities $H_{\text{inter-Kerr},1}\ket{\psi} = t_1'\sum_j b_j^3\ket{j,A}+a^3_j\ket{j,B}$ are chiral symmetric. For the same reason, the nonlinearities  $H_{\text{inter-Kerr},2}\ket{\psi} = t_2'\sum_j b_j^3\ket{j+1,A}+a^3_j\ket{j-1,B}$ are also chiral. However, $H_{\text{inter-Kerr},1}$ reinforces the \textit{intra}-cell coupling $\ket{j,A}\bra{j,B}$ while $H_{\text{inter-Kerr},2}$ reinforces the \textit{inter}-cell coupling $\ket{j+1,A}\bra{j,B}$. Those nonlinearities appear for example in photonic \cite{SelfInducChirTransi}, electrical systems \cite{hadad2018self} and even in phononic devices under some approximations \cite{PhononTopo}. We thus consider a finite SSH chain with such chiral nonlinearities 
\begin{equation}
\left\{ 
\begin{aligned}
    &i \partial_ta_j = (t_1+t_1' |b_j|^2)b_j + (t_2+t_2'|b_{j-1}|^2)b_{j-1}\\
    &i \partial_tb_j = (t_1+t_1' |a_j|^2)a_j + (t_2+t_2'|a_{j+1}|^2)a_{j+1} \ .
\end{aligned}  \right. 
\label{eq:SSHchiral}
\end{equation}
At small amplitude, the linearisation of \eqref{eq:SSHchiral}  yields the usual SSH model, and we find $\mathcal{I}(j)=1$ for $|t_1|<|t_2|$ and $j$ far from the edges. Thus, we predict the existence of a family of chiral nonlinear edge modes $\ket{\psi_s}$ localised on the left A-sites of the lattice (a similar argument would also predicts the existence of non-linear edge modes localised on the right B-sites). This is confirmed by our numerical integration of \eqref{eq:Masterequationchiral} for the model \eqref{eq:SSHchiral} with various choices of parameters $(t_1,t_2,t_1',t_2')$ (the first row of figure \ref{figurechirale}). Interestingly, depending on the competition between inter-cell and intra-cell nonlinear couplings, we find very different behaviours:  When $|t_1'|>|t_2'|$, the amplitude of the edge mode saturates, and the mode becomes a domain wall which invades progressively the bulk. Such a phenomenon was noticed in simulations \cite{PhononTopo} and an experimental setup \cite{Chen13004}, both in mechanical lattices. We unveil here the key hidden role of the generalized chiral symmetry to achieve such a nonlinear topological mode. However, this is not the only possible behavior constrained by chiral symmetry. Indeed, when $|t_1'|<|t_2'|$, we find in contrast that the edge mode remains localized at the boundary, with an increasing amplitude concentrated almost on a single site. For the critical value $|t_1'|=|t_2'|$, the edge mode invades the bulk as in the first case, but with a shape that never saturates. Note that these different behaviors as $s$ varies can in principle be probed experimentally by forcing or pumping the system.
\begin{figure*}[ht]
    \centering
\includegraphics[height=4cm,width = 17cm]{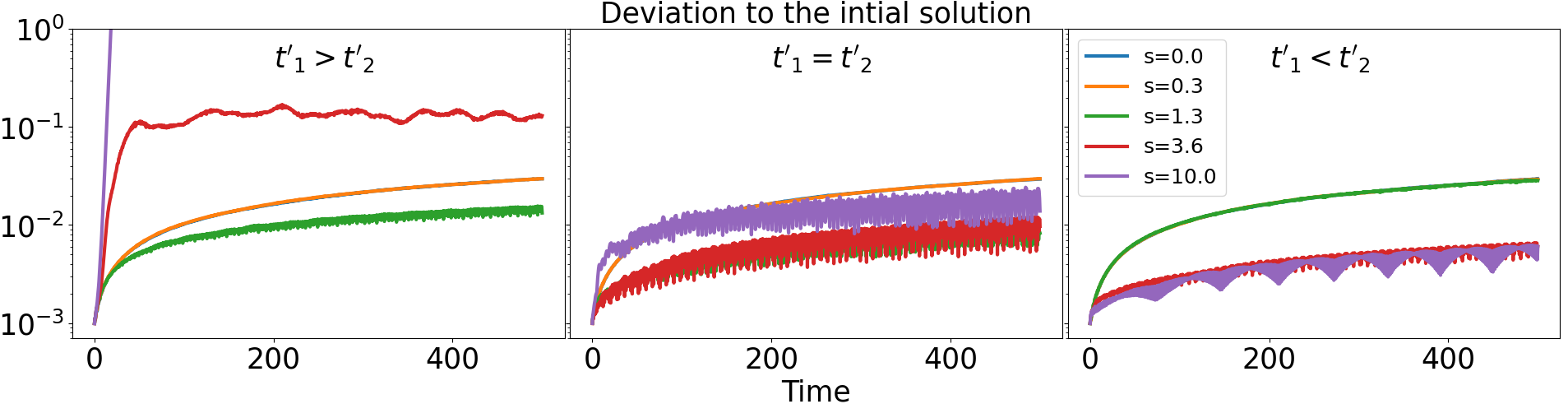}
    \caption{
    Evolution of the deviation $\| \ket{\psi(t)}- \ket{\psi_s}\|$ between the stationary edge states shown in Fig. \eqref{figurechirale} and an initially perturbed one by a random vector of norm $10^{-3}$.}
    \label{fig:stable_chiral}
\end{figure*}
The origin of these different scenarios can be understood by recalling that the nonlinear modes $\ket{\psi_s}$ are obtained by adding iteratively the zero-modes $\ket{\partial_s \psi_s}$ of $H_{\text{eff,s}}=\dd H^{BA}_{\psi_s}$ whose locations are themselves accounted by the variation of $I(x)$ (figure \ref{figurechirale}). Since $\dd H^{BA}_\psi$ reads
\begin{equation}
\bra{j,B}\dd H^{BA}_\psi\ket{\delta \psi}_A = t_{1,\text{eff}} \delta a_{j}+t_{2,\text{eff}} \delta a_{j+1} 
 \end{equation}
with $t_{1,\text{eff}} = t_1+3t_1' |a_j|^2$ and $t_{2,\text{eff}} = t_2+3t_2' |a_{j+1}|^2$, then, when the $a_j$'s are small enough, $|t_{1,\text{eff}}|<|t_{2,\text{eff}}|$ so that $\dd H^{BA}_\psi$ is in the topological phase with $\mathcal{I}(j)=1$ in the bulk. But when increasing the amplitude of the $a_j$'s, one may switch to the trivial phase $\mathcal{I}(j)=0$ where $|t_{1,\text{eff}}|>|t_{2,\text{eff}}|$. If one assumes for simplification that $|a_j| \sim |a_{j+1}| \sim a$, it is clear that the system remains topological even in the high amplitude regime provided that $|t_1'|<|t_2'|$. On the contrary, if $|t_1'|>|t_2'|$, the system undergoes a transition toward a trivial regime where  $|t_{1,\text{eff}}|>|t_{2,\text{eff}}|$. Lastly, when $|t_1'|=|t_2'|$, one gets $|t_{1,\text{eff}}|\sim|t_{2,\text{eff}}|$ at high amplitude leading to a gapless system with $0<\mathcal{I}(j)<1$.

As the amplitude $a_j$ actually depends on the position, the system must be though as being divided into two regions separated by some threshold position $j_s$: The region $j>j_s$ where $|t_{1,\text{eff}}|<|t_{2,\text{eff}}|$ corresponding to the topological phase $(I(j)=1)$), and the region $j<j_s$ where $|t_{1,\text{eff}}|>|t_{2,\text{eff}}|$ corresponding to the trivial one $(I(j)=0)$). At the edge of the topological phase, $I(j)$ must interpolates between $0$ and $1$ implying therefore the existence of a zero-mode of $\dd H^{BA}_\psi$ near by.
As long as $|t_1'|<|t_2'|$, a transition toward the trivial region cannot occur, and so the zero-energy mode remains localised near the edge. In contrast, if $|t_1'|>|t_2'|$, the effective boundary $j_s$ shifts when increasing the amplitude and dissociates from the physical boundary of the chain. Since $\ket{\partial_s\psi_s}$ is localised around $j_s$, it shifts toward the bulk while keeping its shape. As a result, $\ket{\psi_s}$ saturates and invades the bulk. The same reasoning applies when $|t_1'|=|t_2'|$, except that $j_s$ becomes an interface  between a topological and a \textit{gapless} phase. As a result, $\ket{\partial_s\psi_s}$ decreases slowly far away from $j_s$ into the gapless region, leading to a profile of $\ket{\psi_s}$ which is neither flat ($|t_1'|>|t_2'|$) nor exponential ($|t_1'|<|t_2'|$). 

Now that we have established the stationary properties of these topological edge modes, we can look for their nonlinear stability against random perturbations as we did in the non-chiral example. Our results are displayed in figure \ref{fig:stable_chiral}. While the energy $E_s$ remains at zero due to the chiral symmetry, we observe that the topological mode is unstable in the case where $|t_1'|>|t_2'|$ about when the plateau of the domain wall is forming, that is from $s \sim 3-4$. In the two other cases however, we find a relative stability of the edge modes with a deviation that remains relatively small (of order $\sim 10^{-2}$) even at large time $t\gg1$.

To sum up, we have investigated the fate of topological edge states in 1D nonlinear lattices, and showed that those eventually disappear at sufficiently large amplitude, unless the nonlinearities satisfy a generalized chiral symmetry. In that case, a local topological index correctly accounts for the existence and the spatial extension of the nonlinear edge modes, whose actual profile depends on the interplay between the nonlinearities and the underlying topology of the family of linearized Hamiltonians. Our theoretical approach lies on the general hypothesis (i), (ii), (iii) and then the chiral condition \eqref{eq:Chiralcondition} under which the systems \eqref{eq:masterequation} and \eqref{eq:Masterequationchiral} can always be constructed. Therefore we expect the stationary behaviors we describe to not qualitatively change as long as those general hypothesis are verified, for instance if one considers nonlinearities in other nonlinear powers than three. On the other side, it is possible that the stability properties of edge modes are more model-dependant. An extension of our approach to higher dimension,  possibly with other symmetries, is a promising perspective in the search of exotic nonlinear topological states.

\appendix

\section{Numerical computations}

In this paper we shown that if we have an edge solution of $i\partial_t \ket{\psi}=H_\psi \ket{\psi}$ then we can find other edge solutions by solving the following system of differential equation:
\begin{equation}
\left(\dd H_\psi-E_s\right) \ket{\partial_s \psi_s} = (\partial_s E_s) \ket{\psi_s} \label{eq:appendix_master_equation}
\end{equation}

The most common way to numerically solve this kind of differential system is by using iterative method like the Runge-Kutta ones. The only problem that we have to deal in order to apply these methods is to give a valid solution $(\ket{\delta \psi},\delta E)$ for each $s$ of following linear system:

\begin{equation}
\left(\dd H_\psi-E_s\right) \ket{\delta \psi} = (\delta E) \ket{\psi_s} 
\end{equation}

The vector-space of solution of this system can in general be determined by numerical algorithm (the simple one being Gaussian elimination). Once this vector space is determined, we then have to choose one solution in it. This choice is, in general, arbitrary and one can use different method to do it. In general it is often better to choose solutions for $(\ket{\partial_su_s},\partial_sE_s)$ which are close of each other for close $s$ as we observe that the Runge-Kutta method is more stable and fast in these cases. 

For our numerical simulation we use the following procedure. First for $s=0$ we choose one of the two solutions of the system where $\ket{\delta u}$ is of norm 1 and real. Then iteratively, for the system at time $s+\delta s$ we pick the solution which is of norm 1 and is the closest to the one we choose at time $s$.

We control the error made by the numerical procedure by measuring the quantity $||E_s\ket{\psi_s}-H_{\psi_s}\ket{\psi_s}||$ (which should be zero if the procedure is exact). This quantity can be made arbitrarily small at the cost of time of computation. In our example we are able to obtain $||E_s\ket{\psi_s}-H_{\psi_s}\ket{\psi_s}||< 10^{-11}$ and thus the edge mode created are exact up to a negligible error.

In the chiral case where we do not solve \eqref{eq:appendix_master_equation} but the system below, the procedure are almost the same. The only difference is that our variables are only the components of $\ket{\delta \psi}_A$ and not $(\ket{\delta \psi},\delta E)$.
\begin{equation}
    H_{\text{eff,s}}^{BA}\ket{\partial_s\psi_s}_A = 0
\end{equation}

\end{document}